\documentclass[runningheads,a4paper]{llncs}

\usepackage{amssymb}
\usepackage{graphicx}

\usepackage{url}
\urldef{\mailsa}\path|{alfred.hofmann, ursula.barth, ingrid.haas, frank.holzwarth,|
\urldef{\mailsb}\path|anna.kramer, leonie.kunz, christine.reiss, nicole.sator,|
\urldef{\mailsc}\path|erika.siebert-cole, peter.strasser, lncs}@springer.com|    
\newcommand{\keywords}[1]{\par\addvspace\baselineskip
\noindent\keywordname\enspace\ignorespaces#1}

\usepackage{color}
\usepackage[caption=false]{subfig}
\usepackage{cite}

\begin{document}

\mainmatter  

\title{Towards Hybrid Intelligence in Journalism: Findings and Lessons Learnt from a Collaborative Analysis of Greek Political Rhetoric by ChatGPT and Humans}

\titlerunning{Towards Hybrid Intelligence in Data Journalism}

%
%
\author{Thanasis Troboukis$^{a}$ \and  Kelly Kiki$^{a}$ \and  Antonis Galanopoulos$^{b}$ \and  Pavlos Sermpezis$^{c}$ \and Stelios Karamanidis$^{c}$ \and Ilias Dimitriadis$^{c}$ \and Athena Vakali$^{c}$}
\authorrunning{T. Troboukis \textit{et al.}}

\institute{ $^{a}$incubator for Media Education and Development (iMEdD), Athens, Greece\\
\{a.troboukis, k.kiki\}@imedd.org\\
$^{b}$School of Political Sciences, Aristotle University of Thessaloniki, Greece\\
antonisgal@gmail.com\\
$^{c}$School of Informatics, Aristotle University of Thessaloniki, Greece\\
\{sermpezis, skaraman, idimitriad\}@csd.auth.gr\\
}

%
%

\toctitle{Lecture Notes in Computer Science}
\tocauthor{Authors' Instructions}
\maketitle

\begin{abstract}
This chapter introduces a research project titled "Analyzing the Political Discourse: A Collaboration Between Humans and Artificial Intelligence", which was initiated in preparation for Greece's 2023 general elections. The project focused on the analysis of political leaders' campaign speeches, employing Artificial Intelligence (AI), in conjunction with an interdisciplinary team comprising journalists, a political scientist, and data scientists. The chapter delves into various aspects of political discourse analysis, including sentiment analysis, polarization, populism, topic detection, and Named Entities Recognition (NER). This experimental study investigates the capabilities of large language model (LLMs), and in particular OpenAI's ChatGPT, for analyzing political speech, evaluates its strengths and weaknesses, and highlights the essential role of human oversight in using AI in journalism projects and potentially other societal sectors.The project stands as an innovative example of human-AI collaboration (known also as “hybrid intelligence”) within the realm of digital humanities, offering valuable insights for future initiatives. 
\keywords{Political Discourse Analysis, Elections, Artificial Intelligence, ChatGPT, Sentiment Analysis, Topic Detection, Polarization, Populism, Data Journalism}
\end{abstract}

\section{Introduction}\label{sec:introduction}

The incorporation of Artificial Intelligence (AI)\footnote{%
While AI and its concepts are widely known, to ensure a common understanding among all readers of this article, we state the following definitions~\cite{beckett2023}:
\begin{itemize}
    \item \textit{Artificial intelligence (AI)} is a collection of ideas, technologies, and techniques that relate to a computer system’s capacity to perform tasks normally requiring human intelligence.
    \item \textit{Generative AI} is a subfield of AI in its own right,that involves the generation of new data, such as text, images, or code, based on a given set of input data. ChatGPT~\cite{chatgpt} is a prominent example of generative AI.
\end{itemize}
} %
into the practice of journalism presents potential advantages~\cite{beckett2023} in terms of improving the efficiency and excellence of news creation and distribution, as well as bolstering trustworthiness~\cite{opdahl2023trustworthy}. The proliferation of 
AI tools and open datasets made available to journalists to incorporate new (data and AI driven) approaches in their projects~\cite{hermida2019data,borges2016unravelling,broussard2019artificial,hassan2022usage,deuze2022imagination}, is a trend that is only expected to increase in the following years~\cite{survey2022}. 

Advanced AI algorithms, including Natural Language Processing (NLP) models and Large Language Models (LLMs), such as ChatGPT~\cite{chatgpt}, have demonstrated competence in automating labor-intensive activities~\cite{diakopolos2023,gilardi2023chatgpt}, including data mining, sentiment analysis~\cite{bang2023multitask,zhu2023can}, basic reporting, annotating misinformation~\cite{hoes2023using,bang2023multitask,sallam2023chatgpt} and hate speech~\cite{huang2023chatgpt}, etc.. The implementation of automation not only accelerates the pace of news dissemination but also presents opportunities for the field of Investigative Journalism~\cite{stray2019making,rogers2017datajournalism}, allowing human journalists to allocate their efforts toward more intricate analytical and investigative endeavors. Moreover, the ability of AI to analyze extensive datasets can provide a unique understanding of public opinion, political trends~\cite{khudabukhsh2021we,naxera2023more,zhang2022would}, and emerging societal concerns. Therefore, this can provide a richer, more nuanced backdrop for journalists to construct their narratives. 

Nevertheless, it is imperative to acknowledge that AI ought to function as a supplementary instrument to human expertise, rather than a substitute, due to its constraints in comprehending contextual nuances and its propensity for innate biases~\cite{cools2023,mcbride2016ethics}. Therefore, although AI has the potential to significantly transform the field of journalism, its successful incorporation necessitates careful monitoring and ethical deliberation.

\subsubsection*{Motivation and Goals:}

Motivated by (i) the emerging capabilities afforded by AI and (ii) the journalistic need to timely report on a political topic, we launched the project “Analyzing the political discourse: a collaboration between humans and Artificial Intelligence”~\cite{imedd-project} in spring 2023, ahead of the general elections in Greece in May of the same year.

The project aimed to achieve two primary objectives:

\textit{First}, to perform a comprehensive examination of political leaders' campaign speeches, with the aim of uncovering qualitative features of their discourse that go beyond the usual sound bite excerpts presented in the news cycle, such as: identify the main themes that political leaders emphasize in their campaign speeches, as well as to analyze the sentiments conveyed. Additionally, the study quantified  the extent to which elements of polarization and populism are present in the leaders' rhetoric.

\textit{Second}, to explore the boundaries of modern technology and underscore the LLMs proficiency in conducting qualitative and comparative research analysis on political discourse. What are the potential, challenges, and limitations of using AI in journalism projects? How can we craft robust and reliable methodologies? Is human supervision still needed and to what extent?

\subsubsection*{Contributions:} Overall, the project was designed as an experimental collaboration between an interdisciplinary team of individuals, including (data) journalists, a political scientist, computer and data scientists, and web developers, along with AI. This “hybrid intelligence” led to the following contributions.

\textbf{\textit{Methodology and tools:}} We collected a large dataset of 171 speeches of the leaders of political parties (Section~\ref{sec:data-collection}). We crafted a methodology based on journalism and political science methods to analyze the characteristics of the political rhetoric, and automated the analysis using ChatGPT (Section~\ref{sec:prompting}), followed by a human validation of the results (Section~\ref{sec:data-validation}); in total, leading to a “hybrid intelligence” approach. Finally, we created an informative online tool with insightful interactive visualizations (Section~\ref{sec:visualization}) useful to journalists, researchers, and the general public.

While our use case (dataset) was on texts in Greek, our methodology is applicable to any language. To further encourage its use by other journalists, we open-source the data, prompting code and visualizations (see details in~\cite{kiki2023}).

\textbf{\textit{Analysis of political rhetoric during the Greek national elections:}} During our project, and while the pre-election period was still ongoing, we published the results of the analysis\footnote{For each speech, the analysis results were made available within one day of its occurrence.} through our online tool (Section~\ref{sec:visualization}), offering thus a continuous monitoring of the political rhetoric during the elections. This enabled also a number of timely political analyses and articles by journalists, political scientists, academics, etc. (available at~\cite{imedd-project}). We discuss the main findings of the political analyses in Section~\ref{sec:findings}.

\textbf{\textit{Evaluation of ChatGPT’s efficiency for journalism projects:}} Our hybrid intelligence approach (Sections~\ref{sec:prompting} and~\ref{sec:data-validation}) enabled us to evaluate to what extent ChatGPT was efficient in answering our journalistic questions. To this end, we provide a detailed analysis of the accuracy results in Section~\ref{sec:accuracy}. 

Our findings show that ChatGPT demonstrated exceptional proficiency in sentiment analysis and Named Entity Recognition (NER) when applied to the talks of political leaders. Furthermore, it demonstrated efficacy in the detection of the topics that were given in the realm of public discourse. Nevertheless, prior to publication, the data necessitated significant human vetting. ChatGPT exhibited notable bias in its ability to identify discourse associated with polarization and populism. 

\vspace{\baselineskip}
\noindent\textbf{Summarizing}, the incorporation of ChatGPT into our operational framework significantly expedited the examination of political leaders' speeches and the dissemination of our research outcomes, occasionally diminishing the full procedure to a little three-hour timeframe. Nevertheless, the presence of inherent biases within the model, particularly towards left-wing rhetoric, necessitated the implementation of stringent human oversight.


Up to now, conversations about AI in journalism have mostly centered on automating reporting tasks or news delivery. However, this study suggests that journalism should broaden its scope to explore AI's potential in other aspects of the field. For example, AI could revolutionize political reporting by offering data-driven insights that might escape human analysis. This would allow journalists to provide their audience with more fact-based and in-depth examinations of complex, often theoretical, issues.

\section{Methodology}\label{sec:methodology}

\begin{figure}[t]
    \centering    
    \includegraphics[width=\linewidth]{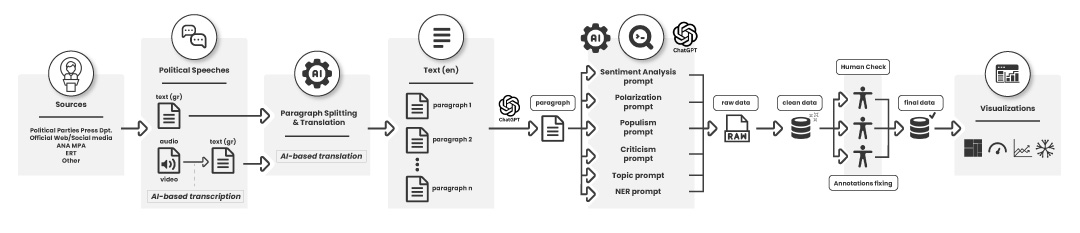}
    \caption{Methodology of the hybrid intelligence political discourse analysis project.}
    \label{fig:methodology}
\end{figure}

In this section, we describe the methodology\footnote{We refer the interested reader to~\cite{kiki2023} for further details of the working methodology.} for collecting the texts of the speeches (Section~\ref{sec:data-collection}), automatically analyzing them using ChatGPT (Section~\ref{sec:prompting}) and then manually validating the results (Section~\ref{sec:data-validation}), and generating the visualizations (Section~~\ref{sec:visualization}) that have been used in our project. Figure~\ref{fig:methodology} depicts the overview of the methodology.

\subsection{Collection and Processing of Speeches}\label{sec:data-collection}

The project's initiation was deliberately synchronized with the commencement of the first election period leading up to the parliamentary elections in Greece on May 21, 2023. Our primary focus was the analysis of campaign speeches delivered by the political leaders of the six parties that held seats in the Greek parliament during the most recent parliamentary term (18th Parliamentary Term, spanning from July 17, 2019, to April 22, 2023). These leaders include:

\begin{itemize}
    \item \textit{Kyriakos Mitsotakis}, Prime Minister and President of New Democracy
    \item \textit{Alexis Tsipras}, President of SYRIZA
    \item \textit{Nikos Androulakis}, President of PASOK-Movement for Change
    \item \textit{Dimitris Koutsoumbas}, Secretary General of the Communist Party of Greece 
    \item \textit{Kyriakos Velopoulos}, President of Greek Solution
    \item \textit{Yanis Varoufakis}, Secretary General of MeRA25
\end{itemize}

Following the May 2023 elections, a so-called “one-day parliament" was formed due to the inability to establish a government in light of the election results. Subsequently, parliamentary elections were called again. In the second phase of our project, during the analysis of campaign speeches leading up to the parliamentary elections on June 25, 2023, we focused on the leaders who were part of the initial study during the first campaign period and represented parties that entered the one-day parliament as a result of the May 2023 elections. This included all of the above leaders except for Y. Varoufakis.

For this study, any public speech delivered by the aforementioned political leaders in a public setting, in the presence of an audience, within the official pre-election period on each occasion, was considered a campaign speech. However, short statements or ad hoc talks (i.e., those lasting less than 15 minutes) and political leaders' interactions with citizens were excluded from the study. Only complete campaign speeches that were available in their entirety were subject to analysis. Due to the fact that K. Velopoulos did not deliver campaign speeches, as defined above, his opening remarks in various press conferences were analyzed. This exception should be noted, as the analysis of Velopoulos' discourse could not be directly compared to that of other political leaders.

To collect the campaign speeches, our primary sources were the press offices of the political parties, followed by the political parties' and their leaders' websites and/or social media accounts. We supplemented our sources list with contributions from the Athens-Macedonian News Agency (ANA-MPA), the Hellenic Broadcasting Corporation (ERT), and other local media outlets in cases where a speech was not fully available through the previously mentioned political parties' communication channels.

The analysis relied on the written text of each political leader's speech whenever available. If the text provided by the parties was already divided into paragraphs based on topics, it was retained in that format. If the written text was not already structured into paragraphs or was organized based on oral delivery, our team of journalists edited the text by segmenting it into paragraphs based on the main topic. In cases where the campaign speeches were not provided in written texts by the parties but were available in audiovisual formats, we used an AI-powered transcription tool\footnote{Transkriptor, \url{https://transkriptor.com/}} to convert the audio into text.

After collecting and/or formatting the written text of each campaign speech in Greek, we automatically translated it into English using a machine translation service\footnote{DeepL, \url{https://www.deepl.com/} through its API.}. Subsequently, ChatGPT was tasked with analyzing this translated speech text, guided by definitions and context provided by the working group.

\subsection{Speech Analysis and Prompting ChatGPT}\label{sec:prompting}

The campaign speeches of political leaders were subjected to analysis at the paragraph level, with the aim of identifying whether the leader primarily focused on criticizing political opponents or presenting their party's agenda, the main topic/theme of the abstract, the dominant sentiment, levels of political polarization and populism detected, and any named entities.

Using an automated way, we prompted ChatGPT (the gpt-3.5-turbo model) 
to provide paragraph-based outputs for the following list of analyses:

\begin{itemize}
    \item A value of either “criticism" or “political agenda" to indicate if the political leader primarily criticized opponents or referred to their party's ideas, opinions, positions, or program proposals.
    \item Identification of the most likely topic/theme discussed from a predefined list of 33 specific topics 
    provided to ChatGPT by the working team (see~\cite{kiki2023} for the detailed list of topics).
    \item Assignment of a sentiment value on a scale of -1 to 1, indicating the negativity, neutrality, or positivity of the passage, to be later classified accordingly based on predefined scales by the working group.
    \item Assignment of a value on a scale of 0 to 1, indicating the level of political polarization within the passage, to be later classified as either no/low, medium, or high polarization based on predefined scales by the working group.
    \item Assignment of a value on the same scale to indicate the level of populism within the passage, to be later classified accordingly.
    \item Extraction of a list of named entities in the passage, categorized as individuals, groups, organizations, political parties, locations, countries, or dates.
\end{itemize}

Some of our prompts to ChatGPT were provided with carefully considered context, aiming to steer the model’s responses by accounting for specific situations that would impact its output, with the goal of mitigating algorithmic bias. For example, when determining whether the text primarily criticized opponents or presented a political agenda and when identifying the main theme, we provided the name and status of the political leader. For instance, “you are reading a passage from a pre-election speech of the Greek Prime Minister, K. Mitsotakis, leader of the New Democracy political party" or “you are reading a passage from a pre-election speech of A. Tsipras, a Greek politician and leader of the opposition party SYRIZA (Coalition of the Radical Left)", was the context provided in cases involving Mitsotakis and Tsipras. To the contrary, we did not disclose the names of the political leaders when evaluating the levels of polarization and populism, in order to bypass ChatGPT's inherent bias, especially towards left-wing rhetoric. Most importantly, when evaluating the levels of polarization and populism in a text, we didn't employ ChatGPT's pre-existing “knowledge"; to mitigate biases stemming from its training data, we provided definitions\footnote{%
We paid high attention to clearly define the two terms and avoid stereotypical approaches and normative bias. The definitions of polarization and populism provided as context to ChatGPT are found in the detailed published methodology~\cite{kiki2023}. For further information on relevant definitions, see~\cite{galanopoulos2023a,galanopoulos2023b}.} %
of polarization and populism as drafted by the political scientist in the working group. This context remained consistent in all cases.

For each unique campaign speech, we compiled a dataset with rows corresponding to the number of paragraphs and columns representing the research variables. Following this, we conducted a human review of the ChatGPT results.

\subsection{The Data Validation Process}\label{sec:data-validation}

For determining whether each speech passage focused on criticizing opponents or presenting a political agenda and identifying the major theme, data validation involved a minimum of two journalists who cross-verified the ChatGPT results. Although the analysis was conducted on text translated into English, validation was performed on the original Greek text. Any inaccuracies were rectified by the working group. Regarding sentiment analysis, polarization, and populism levels, two members of the research team —the political scientist and a journalist— reviewed ChatGPT's outputs. In cases of inaccuracies, the working group made the necessary corrections as follows:

\textit{Sentiment analysis}: Each passage was initially categorized based on ChatGPT's assigned value: negative for scores between -1 and -0.34, neutral for scores between -0.33 and 0.33, or positive for scores between 0.34 and 1. If ChatGPT assigned a sentiment score that placed a paragraph into a different category than human judgment, that value was excluded and treated as missing data in subsequent analysis and relevant visualizations. This approach was adopted to prevent ad hoc human intervention in the sentiment indicator's average, which was used to classify the entire campaign speech as negative, neutral, or positive.

\textit{Levels of polarization and populism}: Each passage was initially categorized based on ChatGPT's assigned value into one of three categories: zero/low level for scores between 0 and 0.5, medium level for scores between 0.51 and 0.8, or high level for scores between 0.81 and 1. If ChatGPT assigned a value that differed from human judgment, the political scientist, a member of the working team, adjusted this value to 0,\footnote{The majority of excerpts in the zero/low level category had scores very close to 0. To avoid the potential introduction of extreme values through human intervention within this category, a corrective value of 0 was established for this specific class. Mean scores were used as the corrective values for the remaining categories.} 0.6, or 0.9, as appropriate for zero/low level, medium level or high level (of either polarization or populism) classification.

\begin{figure}[t]
\centering
\includegraphics[width=0.7\linewidth]{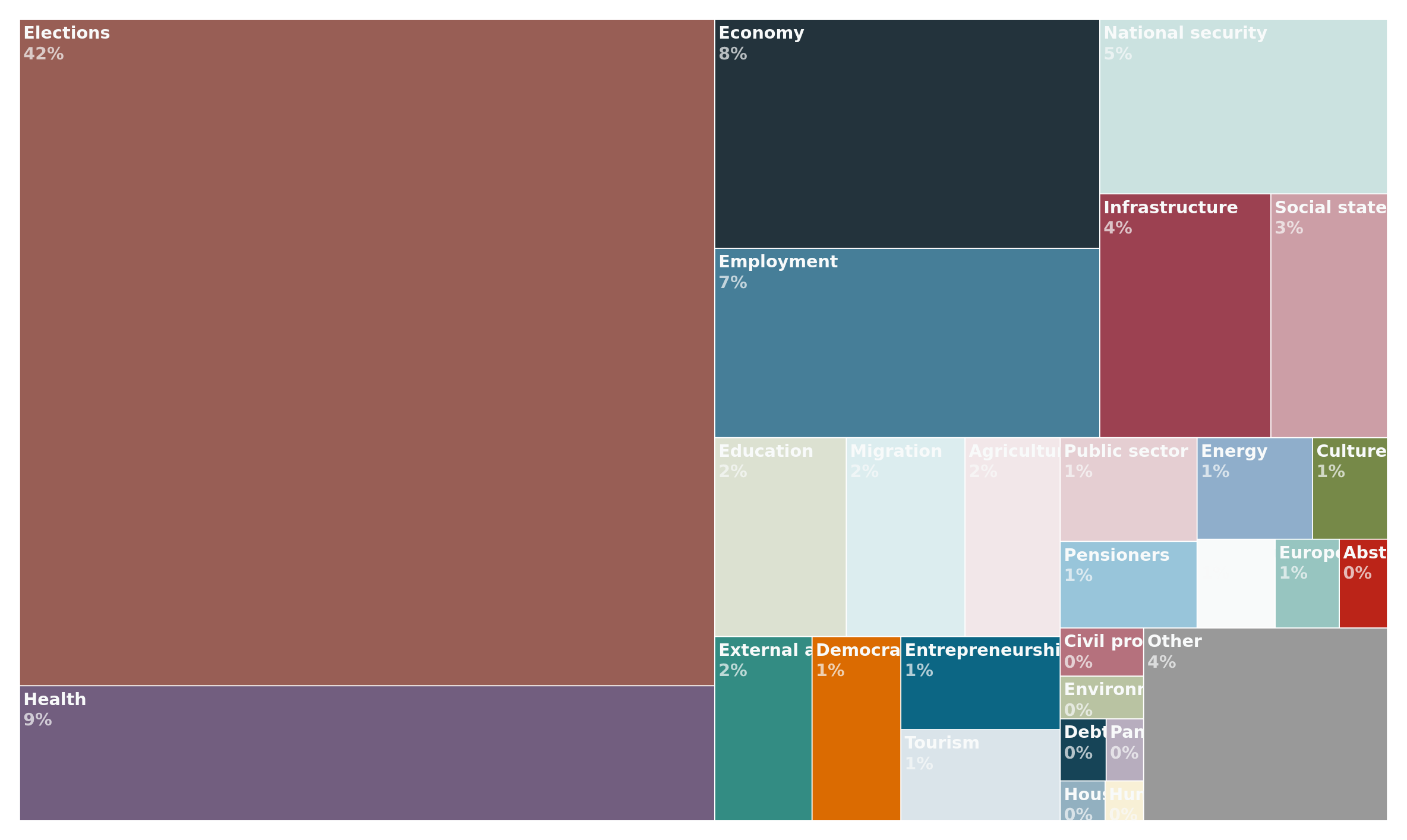}
    \caption{Treemap chart displaying the topics covered in the speeches of K. Mitsotakis (only for the parts that referred to their “political agenda”) during the first election period. Each box corresponds to a different topic, and its size is proportional to the percentage of paragraphs referring to this topic.}
    \label{fig:treemap}
\end{figure}

\subsection{The Data Visualizations}\label{sec:visualization}

The validated results were visualized with various charts 
that were constantly updated as the campaign period evolved. The major visualizations included:

\begin{figure}[t]
\centering
\includegraphics[width=0.6\linewidth]{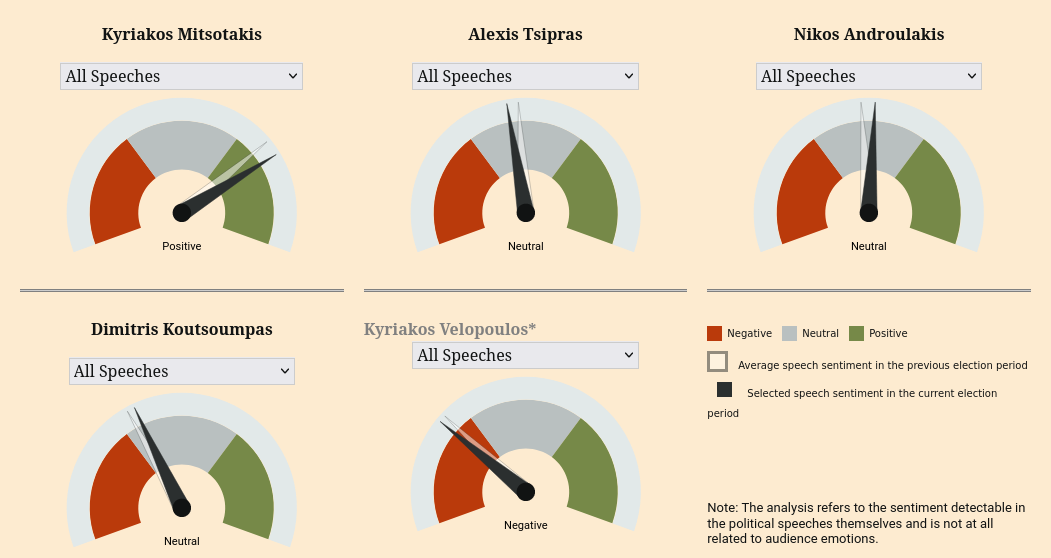}
    \caption{Speedometer charts displaying the average sentiment for each political leader during the first (light gray) and second (dark gray) election periods. Red / gray / green areas correspond to negative / neutral / positive sentiments, respectively.}
    \label{fig:speedometers}
\end{figure}

\begin{figure}[t]
\centering
\includegraphics[width=0.7\linewidth]{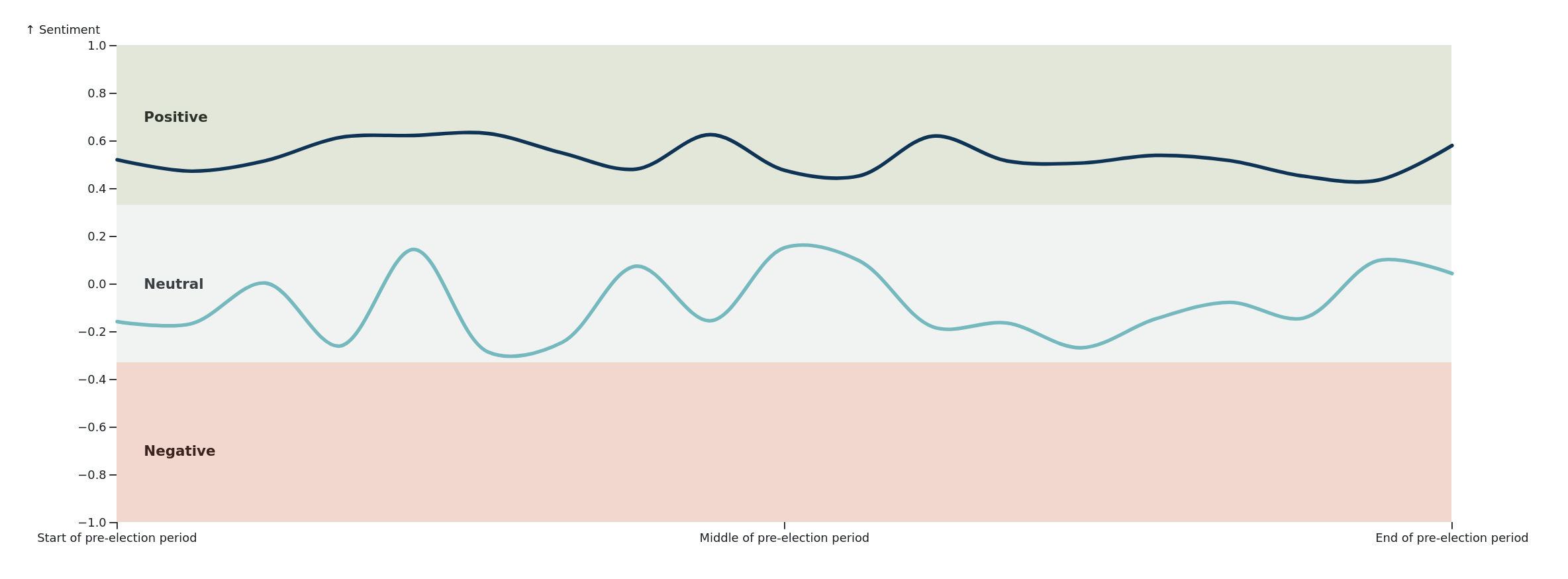}
    \caption{Line chart depicting the sentiment evolution over the course of the second election period (x-axis) for K. Mitsotakis (dark blue line) and A. Tsipras (light blue line). The y-axis corresponds to sentiment values, from -1 to 1, and colored areas to negative (red), neutral (gray) and positive (green) sentiments.}
    \label{fig:sentiment-line}
\end{figure}

\begin{figure}[t]
\centering
    \includegraphics[width=0.7\linewidth]{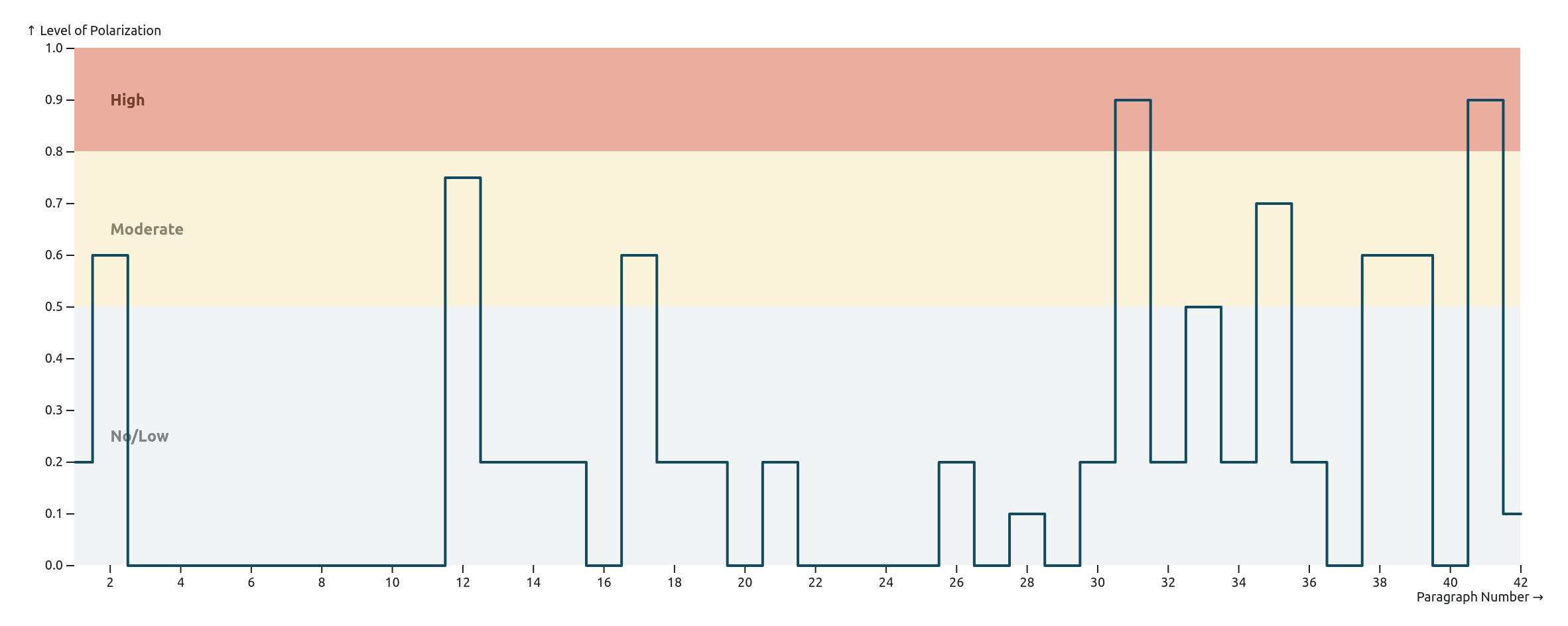}
    \caption{Step chart depicting the evolution of the polarization during a speech of A. Tsipras. The y-axis corresponds to polarization values, from 0 to 1, and colored areas to None/Low (gray), Medium (yellow) and High (red) levels of polarization sentiments.}
    \label{fig:polarization}
\end{figure}

\begin{itemize}
    \item A treemap displaying the shares of topics covered by each political leader in their speeches. These percentages represented the estimated proportion of words spoken on each topic relative to the total number of words in the speech. An example is given in Figure~\ref{fig:treemap}.
    \item A speedometer plotting the sentiment analysis of each speech (Figure~\ref{fig:speedometers}). The position of the needle indicated the average sentiment values at the paragraph level, leading to the classification of the speech as negative, neutral, or positive.
    \item Line charts depicting the sentiment evolution in specific speeches or over the course of the election period for two different leaders, based on the user's selection (Figure~\ref{fig:sentiment-line}).
    \item Step charts illustrating the evolution of polarization and populism levels during each speech or throughout a campaign period (Figure~\ref{fig:polarization}).
    \item A heatmap representing the sentiment by theme discussed in each leader's speeches throughout the campaign period.
    \item A radial dendrogram showing the entities named by each political leader, organized by type.
    \item A bar plot counting the number of times “will" had been mentioned by each political leader across the campaign period. 
\end{itemize}

\section{Findings and Insights}\label{sec:findings}

Understandably, the volume of data resulting from such an analysis of 171 pre-election speeches is particularly large. A detailed description of the findings would be impossible in the limited scope of a chapter and for this reason in this paper we will focus on some indicative findings that highlight the usefulness of the chosen methodology and the synergy of human expertise and AI tools. 
Certainly, the first and general advantage is related to the speed of data processing. As we will see below, the speed of processing, the constant and timely updating of our publications, and the visualization charts (see Section 2.4) that offer a quick aggregate view of the analysis, allowed us to conduct  analyses regarding the agenda and the style of each politician and draw preliminary conclusions in a shorter time frame.\footnote{To be sure, quantitative and in depth analysis is still needed and generally requires time. Our work does not substitute in depth academic analyses but strives to provide tools to support them.}

\subsubsection*{The (unexpected) absence of polarization and populism. }

According to initial expectations at the beginning of the pre-election period, the country would be faced with a particularly polarized political landscape marked, at the same time, by high levels of populism. This estimation, reinforced by the certainty of the two successive elections, leads us to empirically explore this hypothesis by incorporating indicators~\cite{galanopoulos2023a,galanopoulos2023b} of polarization and populism into the analysis. Eventually, our findings did not confirm these assessments. On the first hand, no extreme levels of polarization or strategic and consistent use of polarizing discursive patterns was observed in the speeches of the political leaders: the average populism and polarization scores over all speeches were 0.2 and 0.1, respectively, on a scale from 0 to 1. Nevertheless, it is important to note that there were occasional instances of polarization during both election periods.\footnote{Notably, D. Koutsoumpas, general secretary of the communist party, recorded the highest escalation in the polarization index during the second election period (average score of 0.24).} On the other hand, populism was not a prominent discursive strategy employed by any political leader. More frequent uses of populist discursive patterns could be found in A. Tsipras' speeches during the first election period, although they remained at fairly low levels (slightly higher than 0.1), while in the second the use of populist rhetoric was almost completely absent.\footnote{A. Tsipras had articulated a characteristically populist discourse in the past, especially during 2012-2015~\cite{stavrakakis2014left}. Our findings confirm that populism is not an omnipresent feature of a party or a politician. Instead, populist discourse presents fluctuations over time. To properly identify a party as a populist or not, we need to study its official discourse at any given conjuncture. }

\subsubsection*{What (topics) and how (sentiment) politicians talked about.}

An important advantage of our project was the fact that, by identifying and categorizing the primary topics discussed by political leaders (e.g., Figure~\ref{fig:treemap}), we were able to assess their policy priorities. More specifically, our ability to discern which issues were high or low on each party's agenda was greatly enhanced by the use of AI. Comparing the political style and strategy of each leader, our analysis points to pertinent remarks regarding political communication during a campaign period. K. Mitsotakis’ discourse demonstrated impressive consistency, characterized by a positive sentiment (as demonstrated in Figure~\ref{fig:sentiment-line}) and a dedicated focus on promoting the party’s programmatic agenda (more than 85\% of his speeches). Moreover, it is worth noting that the thematic agenda around the topics of \textit{healthcare} (9\%), \textit{economy} (8\%), and \textit{employment} (7\%) championed by New Democracy shaped to a great extent the overall debate. This positive and programmatic dimension, which are closely intertwined, saw a slight increase in emphasis during the second election period, likely influenced by the first positive outcome. Finally, Mitsotakis refrained from mentioning his political opponents by name. 

Regarding the other political leaders (Figure~\ref{fig:speedometers}), A. Tsipras and N. Androulakis maintained a distinctively neutral sentiment in their speeches, whereas D. Koutsoumpas and Y. Varoufakis leaned more towards neutrality but occasionally displayed a negative sentiment. In contrast, K. Velopoulos consistently conveyed a negative sentiment in his speeches. During the second election period, A. Tsipras’ discourse underwent two notable changes compared to the first. Firstly, there was a significant reduction in populist discursive patterns, as it was described earlier. Secondly, there was a systematic effort to place greater emphasis on the programmatic agenda of the party (from 50\% in the first elections to 54\% in the second). In terms of the topics discussed, the political debate primarily revolved around the economy. However, in the second period, Tsipras also sought to highlight the importance of healthcare (18\%), which became a central focus of his criticism towards New Democracy. Additionally, he addressed issues of accountability, corruption, and democracy. On the other hand, certain important topics such as the environment, culture, energy, agricultural policy, infrastructure, national security, and migration received less attention (less than 5\% each for most politicians).

\subsubsection*{Emerging topics during the two pre-election periods: the case of migration.}

Finally, as we argued earlier in depth analysis using various methodologies can and should be carried out regarding the two recent elections in Greece. Our project can offer different points of entrance and data for the formulation of research hypotheses, especially in the fields of agenda setting, discourse analysis and framing processes. Using our multi-faceted approach, we managed to identify very soon that the main thematic change in the agenda and the speeches of every politician during the second period was their stance on the migration issue.
The devastating sinking of a fishing boat carrying hundreds of refugees and migrants off Pylos on Wednesday, June 14, 2023,~\cite{piper2023} brought the issue of migration and refugees to the forefront of public debate. Until the aforementioned tragedy occurred, the issue of migration was either completely absent from the agenda of political leaders or received only passing mentions in their speeches. Of particular interest is how the issue was framed and the connotations associated with it. Indeed, a strikingly contrasting picture emerges when examining the election speeches of the leaders of the two largest parties, New Democracy and SYRIZA.
Although migration was not a central theme of K. Mitsotakis’ campaign speeches, he mentioned it more frequently than any other political leader. In nearly half of his speeches during the first election campaign, there is at least one mention or passing reference to the immigration issue. These references were present in the second election period, both before and after the deadly shipwreck. However, it cannot be claimed that immigration is a top priority on K. Mitsotakis’ agenda, as an overall analysis of his election speeches reveals that migration constitutes only 2\% of his programmatic discourse. These references primarily emphasize a specific aspect: border protection. K. Mitsotakis’ overall approach to migration in his election speeches revolves around this perspective and the issue was framed as exclusively a matter of national security. On the other hand, the migration issue was completely absent from A. Tsipras' campaign speeches up until the last week of June elections.\footnote{It is important to underline that only the leaders’ election speeches were analyzed (party events, interviews, statements, articles, etc. were not taken into account).}

\section{Effectiveness and Accuracy of ChatGPT}\label{sec:accuracy}

As previously outlined, the analysis of the political speech was conducted on a per-paragraph basis. For each paragraph, ChatGPT assigned values to various metrics and categories, which are detailed in Table~\ref{tab:chatgpt-accuracy}. Human annotators made corrections to the categories assigned by ChatGPT when they believed that a different category better matched the content of that part of the speech.\footnote{Only for the “sentiment” metric we do not correct, but just annotate as incorrect when ChatGPT places a paragraph into a different category than human judgment. }

To assess ChatGPT's accuracy in comparison to human annotations, we calculate it as the \textit{fraction of paragraphs where the ChatGPT annotation remains unchanged by human annotators, divided by the total number of paragraphs}.

The accuracy results, presented as percentages, can be found in Table~\ref{tab:chatgpt-accuracy}. In the ideal scenario, where ChatGPT's annotations align perfectly with those of human annotators, the accuracy would be 100\%. Additionally, we include the frequency of the most commonly occurring category for each metric. This frequency represents the accuracy of a “dummy" model and serves as a baseline for comparing the accuracy of ChatGPT's results.

\begin{table}[h]
    \centering
    \caption{Accuracy of ChatGPT annotations (with respect to manual annotation of humans) per metric.}
    \label{tab:chatgpt-accuracy}
    \begin{tabular}{l|l|r}
    Task & Most prevalent category (\%) & Accuracy\\
    \hline
    sentiment & Positive (40\%) & 94.2\%\\
    topics & Elections (26\%) & 61.7\%\\
    polarization & Zero/Low (86\%) & 87.4\%\\
    populism & Zero/Low (96\%) & 89.8\%\\
    criticism or agenda & political agenda (61\%) & 89.3\%
    \end{tabular}
\end{table}

\textbf{Sentiment}: In the task of sentiment detection, ChatGPT attains a notably high accuracy rate of 94.2\%, the highest among all tasks. This suggests its robust suitability for sentiment analysis in political speeches, a conclusion aligned with recent findings in the field~\cite{zhu2023can,bang2023multitask}.

\textbf{Criticism vs political agenda}: ChatGPT also exhibits a remarkable ability to differentiate between politicians critiquing their opponents and presenting their own political agenda in nearly 9 out of 10 instances.

\textbf{Polarization and populism}: The accuracy metric for detecting speech containing elements of polarization or populism are high as well. However, it is important to note that in both cases, the majority of paragraphs were annotated -by an expert political scientist- as lacking these characteristics (see “Zero/Low" percentages in Table~\ref{tab:chatgpt-accuracy}). A dummy classifier consistently labeling text as “Zero/Low" for polarization or populism would achieve similar accuracy. These results suggest that ChatGPT may not be suitable for unsupervised detection of such specific political speech concepts. In fact, ChatGPT tends to overestimate the level of polarization/populism in text, providing results more aligned with a broad interpretation of these concepts (as opposed to the specific definitions in political science, as outlined in Section~\ref{sec:findings}). This overestimation may result from biases or inaccuracies present in the training data used for ChatGPT.

\textbf{Topics}: ChatGPT assigns a topic from a predefined list of 33 topics to a paragraph, aligning with human annotation in 61.7\% of cases. While this accuracy rate may seem relatively low, it is essential to consider that this task is inherently challenging (i.e., multiclass classification with 33 classes), and ChatGPT outperforms a dummy model by nearly threefold (26\%). To further validate this, we trained a state-of-the-art NLP model~\cite{roberta2019} on our dataset and evaluated its accuracy. Even though it was trained on the same dataset and used human labels (an opportunity that was not available during our project's inception), it achieved only slightly higher accuracy than ChatGPT.

\section{Conclusions}\label{sec:conclusions}

The interdisciplinary nature of our team, composed of individuals with diverse backgrounds within the working group, prompts us to emphasize that, based on our experience with this project, journalism could explore and delve into specific areas, political science could focus on others, and computer science could comment on different, additional aspects. This does not imply that there are no shared findings; instead, it underscores that experts from each field may approach these shared findings differently based on their respective criteria and expertise.

Overall, as expected, the major advantage of integrating AI into this work was the elimination of the initial data processing and analysis speed bottleneck. This comes with additional benefits: First, it allows journalism to provide constantly and timely updated pieces, which can be published shortly after the speeches are delivered. Second, it enhances the journalistic content itself by facilitating timely comparative analysis of the agenda and rhetorical style of each politician. Third, it enables the review of initial assessments early on, shedding light on positions influenced by public sentiment or narrative when they do not align with scientific criteria. This was evident when we discovered that there were low levels of polarization/populism, contrary to what was discussed in the public~sphere.

Furthermore, a significant advantage of this type of thematic analysis is that the identified topics in politicians' public speeches allow researchers and the general public to assess their policy priorities based on their entire body of discourse. Alternatively, one can focus on specific speeches to evaluate how these communicated priorities might vary depending on the local audience. Having a continuously updated set of topics that political leaders typically raise can be a valuable resource for revisiting and examining their positions over time and whether their priorities have changed in response to significant events that may influence political rhetoric. This was evident in our project when analyzing their stances on migration issues before and after the tragic sinking incident.

On the other hand, when examining the effectiveness and accuracy rates of ChatGPT, it becomes evident that while it is certainly suitable for specific tasks, such as sentiment analysis, its overall success may not be consistently reliable in all situations. In the event of redoing the topic analysis, we would adopt an alternative methodology. Namely, we would prompt ChatGPT to autonomously discern each paragraph's prevailing theme, followed by classifying these themes into one of our 33 pre-defined topics. In any case, the results obtained from the topic detection task reveal a thought-provoking aspect of using AI in journalism. While ChatGPT can achieve state-of-the-art accuracy, it may still fall short of meeting journalistic standards, particularly when dealing with sensitive subjects like political reporting. 


This highlights the necessity for human oversight and underscores the importance and requirement for human-in-the-loop AI processes and tools in journalism. This need for human oversight is also apparent in the process of prompting and providing context, which is aimed at reducing bias and minimizing the inclusion of outdated information in ChatGPT’s responses.

Nonetheless, the recent opportunity to train an interactive AI model based on custom data is a promising development for future studies. It also has the potential to shift human involvement from verification to analysis.

\bibliographystyle{splncs04} 

\begin{thebibliography}{10}
\providecommand{\url}[1]{\texttt{#1}}
\providecommand{\urlprefix}{URL }
\providecommand{\doi}[1]{https://doi.org/#1}

\bibitem{bang2023multitask}
Bang, Y., Cahyawijaya, S., Lee, N., Dai, W., Su, D., Wilie, B., Lovenia, H., Ji, Z., Yu, T., Chung, W., et~al.: A multitask, multilingual, multimodal evaluation of chatgpt on reasoning, hallucination, and interactivity. arXiv:2302.04023  (2023)

\bibitem{beckett2023}
Beckett, C., Yaseen, M.: Generating Change: A Global Survey Of What News Organisations Are Doing With Artificial Intelligence. JournalismAI, LSE (2023)

\bibitem{borges2016unravelling}
Borges-Rey, E.: Unravelling data journalism: A study of data journalism practice in british newsrooms. Journalism Practice  \textbf{10}(7),  833--843 (2016)

\bibitem{broussard2019artificial}
Broussard, M., Diakopoulos, N., Guzman, A.L., Abebe, R., Dupagne, M., Chuan, C.H.: Artificial intelligence and journalism. Journalism \& mass communication quarterly  \textbf{96}(3),  673--695 (2019)

\bibitem{cools2023}
Cools, H., Diakopoulos, N.: Towards guidelines for guidelines on the use of generative ai in newsrooms (2023), \url{https://generative-ai-newsroom.com/towards-guidelines-for-guidelines-on-the-use-of-generative-ai-in-newsrooms-55b0c2c1d960}

\bibitem{survey2022}
{Data Journalism}: Survey – state of data journalism 2022 (2022), \url{https://datajournalism.com/survey/2022/}

\bibitem{deuze2022imagination}
Deuze, M., Beckett, C.: Imagination, algorithms and news: Developing ai literacy for journalism. Digital Journalism  \textbf{10}(10),  1913--1918 (2022)

\bibitem{diakopolos2023}
Diakopoulos, N.: The state of ai in media: From hype to reality (2023), \url{https://generative-ai-newsroom.com/the-state-of-ai-in-media-from-hype-to-reality-37b250541752}

\bibitem{galanopoulos2023b}
Galanopoulos, A.: Detecting polarization in the pre-election political discourse in greece (2023), \url{https://lab.imedd.org/en/polarization-greek-political-speech/}, iMEdD Lab

\bibitem{galanopoulos2023a}
Galanopoulos, A.: Populism in pre-election political discourse in greece (2023), \url{https://lab.imedd.org/en/populism-in-pre-election-political-discourse-in-greece/}, iMEdD Lab

\bibitem{gilardi2023chatgpt}
Gilardi, F., Alizadeh, M., Kubli, M.: Chatgpt outperforms crowd-workers for text-annotation tasks. arXiv preprint arXiv:2303.15056  (2023)

\bibitem{hassan2022usage}
Hassan, A., Albayari, A.: The usage of artificial intelligence in journalism. Future of Organizations and Work After the 4th Industrial Revolution p.~175

\bibitem{hermida2019data}
Hermida, A., Young, M.L.: Data journalism and the regeneration of news. Routledge (2019)

\bibitem{hoes2023using}
Hoes, E., Altay, S., Bermeo, J.: Using chatgpt to fight misinformation: Chatgpt nails 72\% of 12,000 verified claims  (2023)

\bibitem{huang2023chatgpt}
Huang, F., Kwak, H., An, J.: Is chatgpt better than human annotators? potential and limitations of chatgpt in explaining implicit hate speech. arXiv:2302.07736  (2023)

\bibitem{imedd-project}
{iMEdD}: A special edition of {iMEdD} for the general elections in greece (2023), \url{https://lab.imedd.org/en/elections-2023/}

\bibitem{khudabukhsh2021we}
KhudaBukhsh, A.R., Sarkar, R., Kamlet, M.S., Mitchell, T.: We don't speak the same language: Interpreting polarization through machine translation. In: Proc. of the AAAI Conference on Artificial Intelligence. vol.~35, pp. 14893--14901 (2021)

\bibitem{kiki2023}
Kiki, K., Troboukis, T., Galanopoulos, A., Sermpezis, P., Karamanidis, S., Dimitriadis, I.: How we analyze the campaign speeches of political leaders (2023), \url{https://lab.imedd.org/en/pos-analyoume-tis-proeklogikes-omilies-ton-politikon-archigon/}

\bibitem{roberta2019}
Liu, Y., Ott, M., Goyal, N., Du, J., Joshi, M., Chen, D., Levy, O., Lewis, M., Zettlemoyer, L., Stoyanov, V.: Roberta: A robustly optimized bert pretraining approach. arXiv:1907.11692  (2019)

\bibitem{mcbride2016ethics}
McBride, R.E.: The ethics of data journalism  (2016)

\bibitem{naxera2023more}
Naxera, V., Ka{\v{s}}e, V., Stul{\'\i}k, O.: ‘the more populism types you know, the better political scientist you are?’machine-learning based meta-analysis of populism types in the political science literature. Jnl. of Contemporary European Studies  (2023)

\bibitem{opdahl2023trustworthy}
Opdahl, A.L., Tessem, B., Dang-Nguyen, D.T., Motta, E., Setty, V., Throndsen, E., Tverberg, A., Trattner, C.: Trustworthy journalism through ai. Data \& Knowledge Engineering  \textbf{146},  102182 (2023)

\bibitem{chatgpt}
OpenAI: Chatgpt: Conversational ai service (2023), \url{https://openai.com/blog/chatgpt}

\bibitem{piper2023}
Piper, I., Lee, J., Parker, C., Labropoulou, E.: Tracing a tragedy: How hundreds of migrants drowned on greece’s watch (2023), \url{https://www.washingtonpost.com/world/interactive/2023/greece-migrant-boat-coast-guard/}

\bibitem{rogers2017datajournalism}
Rogers, S., Schwabish, J., Bowers, D.: Data journalism in 2017. Google News Lab  (2017)

\bibitem{sallam2023chatgpt}
Sallam, M., Salim, N.A., Ala’a, B., Barakat, M., Fayyad, D., Hallit, S., Harapan, H., Hallit, R., Mahafzah, A., Ala'a, B.: Chatgpt output regarding compulsory vaccination and covid-19 vaccine conspiracy: A descriptive study at the outset of a paradigm shift in online search for information. Cureus  \textbf{15}(2) (2023)

\bibitem{stavrakakis2014left}
Stavrakakis, Y., Katsambekis, G.: Left-wing populism in the european periphery: the case of {SYRIZA}. Journal of political ideologies  \textbf{19}(2),  119--142 (2014)

\bibitem{stray2019making}
Stray, J.: Making artificial intelligence work for investigative journalism. Digital journalism  \textbf{7}(8),  1076--1097 (2019)

\bibitem{zhang2022would}
Zhang, B., Ding, D., Jing, L.: How would stance detection techniques evolve after the launch of chatgpt? arXiv:2212.14548  (2022)

\bibitem{zhu2023can}
Zhu, Y., Zhang, P., Haq, E.U., Hui, P., Tyson, G.: Can chatgpt reproduce human-generated labels? a study of social computing tasks. arXiv:2304.10145  (2023)

\end{thebibliography}

\end{document}